\documentstyle[aps,preprint]{revtex}
\begin{document}

\title{Size and Scaling
in Ideal Polymer Networks}

\author{Michael P. Solf,
Thomas A. Vilgis\footnote{E-mail: vilgis@mpip-mainz.mpg.de}\footnote{to
whom all correspondence should be send}}

\address{Max-Planck-Institut f\"ur Polymerforschung,
Postfach 3148, 55021 Mainz, Germany} 

\date{\today} 
\maketitle
\begin{abstract}
The scattering function and radius of gyration of an ideal
polymer network are calculated depending on the strength of
the bonds that form the crosslinks.  Our calculations are
based on an {\it exact} theorem for the characteristic
function of a polydisperse phantom  network that allows for
treating the crosslinks between pairs of randomly selected
monomers as quenched variables without resorting to replica
methods.  From this new approach it is found that the
scattering function  of an ideal network obeys a master
curve which depends on one single parameter $x=
(ak)^2 N/M$, where $ak$ is the product of the
persistence length times the scattering wavevector, $N$ the
total number of monomers and $M$ the crosslinks in the
system.  By varying the crosslinking potential from
infinity (hard $\delta$-constraints) to zero (free chain),
we have also studied the crossover of the radius of
gyration from the collapsed regime where $R_{\mbox{\tiny
g}}\simeq {\cal O}(1)$ to the extended regime
$R_{\mbox{\tiny g}}\simeq {\cal O}(\sqrt{N})$. In the
crossover regime the network size $R_{\mbox{\tiny g}}$ is
found to be proportional to $(N/M)^{1/4}$. The latter
result can  be understood in terms of a simple Flory
argument.
\end{abstract}

\pacs{61.41.+e, 64.60.Cn, 87.15.By}
\maketitle


\section{Introduction}

Sufficiently crosslinked macromolecules form solid-like
rubber networks whose spectacular elastic properties are
commonly believed to be of entropic origin.  As a
consequence polymeric networks are often successfully
modeled as a set of independent random walks with the only
restriction that the end points deform affinely under
external stress. In a real network, however, permanent
junctions between the macromolecules are randomly formed
upon fabrication leading to a high degree of
polydispersity. To meet with this complicated physical
situation more sophisticated theoretical models are
required.

Recently various analytical studies
\cite{deamed,golgol,vilsol,pany,hiba,waed,edvil} have aimed at a
statistical description of polymer networks taking
randomness of the crosslinking positions into account. The
mathematical challenge that arises in any of these
approaches is that the permanent junction-points between
macromolecules are frozen and cannot be treated within the
framework of Gibbsian statistical mechanics.  This has
first been realized by Edwards \cite{deamed} and his
replica formalism has by now become the standard approach
in the field of polymer networks.  Unfortunately the
replica method forces strong approximations to remain
analytically tractable. A recent work by Panyukov and Rabin
\cite{pany} seems to overcome several difficulties and
derives promising results for the elasticity.

In this study we use a different route of thought. It has
recently been shown \cite{solvil} that for randomly
crosslinked Gaussian structures substantial progress can be
made by invoking quite different mathematical tools than
replica field theory. The purpose of the paper is to report
on further progress in this direction. Our working model is
an ideal polymer chain of $N$ monomers with $M$ randomly
selected pairs of monomers constrained to be in
close neighborhood. The distance constraints are modelled
by harmonic potentials. By varying the strength of the
crosslinking potential we can continuously switch from a
network situation with hard $\delta$-constraints to the
case of a free chain.

Although the above model is highly idealized, the system is
most interesting in its own right, since it is considered a
first step towards a systematic theory of polymer networks
and gels. Moreover, the statistics of a huge macromolecule
crosslinked to itself (randomly or not) has attracted a lot
of attention recently because of its possible implications
to protein structure reconstruction from NMR data
\cite{gutin,bry,kant}. Despite of the recent interest 
to the best of our knowledge these are the first exact
results for ideal networks that have been reported.

\section{The ideal network}

We adopt the minimal model of a huge Gaussian chain that is
$M$ times crosslinked to itself. In the Hamiltonian only
terms that model chain connectedness and contributions due
to crosslinking are retained.  Complicating factors such as
entanglements, excluded volume are deliberately neglected
from the start. An appropriate Hamiltonian to begin with is
\begin{equation}
\label{1}
\beta {\cal H}_0=\frac d{2a^2}\sum_{i=1}^N({\bf R}_i-{\bf
R}_{i-1})^2+\frac {d}{2\varepsilon ^2}\sum_{e=1}^M({\bf
R}_{i_e}-{\bf R}_{j_e})^2~.
\end{equation}
We have assumed $N+1$ monomers whose locations in space are
given by $d$-dimensional vectors ${\bf R}_i$
($i=0,1,...,N$).  Distance constraints exist between pairs
of monomers labeled by $i_e$ and $j_e$. For further use we
introduce the inverse strength of the crosslinking
potential
\begin{equation}
\label{2}
z=\left(\frac{\varepsilon}{a}\right )^2
\end{equation}
as the mean squared distance between monomers that form the
crosslinks measured in units of the persistence length $a$
of the chain (figure 1).  Limiting cases are given by
$z=0$ (hard ${\delta}$-constraints) and $z\rightarrow
\infty$ (free chain). The whole crosslinking topology is
specified by a set of $2M$ integers
C=$\{i_e,j_e\}_{e=1}^M$. It has been shown
\cite{solvil} that the model in (\ref{1}) is equivalent to the
Deam-Edwards model \cite{deamed} without excluded-volume
interaction if averages are understood in the following
sense
\begin{equation}
\label{3}
\Big\langle ....\Big\rangle_0= \lim_{z \rightarrow 0}\frac{
{\displaystyle\int} \prod_{i=0}^N d{\bf R}_i\,e^{-\beta
{\cal H}_0}....}{{\displaystyle \int} \prod_{i=0}^Nd{\bf
R}_i\,e^{-\beta {\cal H}_0}}~.
\end{equation}
It is interesting to note that the entire range $0 \geq z < \infty$
can be treated with the same formalism. In \cite{solvil} we have  
explicitly
shown that all limits are mathematically well defined. Thus the  
here presented
method is indeed able to cover the range of all crosslink  
strengths, i.e.,
from very soft to the quenched case. 

To model $M$ {\it uncorrelated} crosslinks the distribution
of frozen variables C is assumed to be uniform
\begin{equation}
\label{4}
\prod_{e=1}^M \bigg\{ \frac{2}{N^2}
\sum_{0\leq i_e<j_e\leq N}
\bigg \}
\end{equation}
Other distributions are in principle possible but not
considered in this investigation. As usual for systems with
permanent constraints care must be taken in evaluating
averages of physical quantities. The strategy here is not
to start with the quenched average over the frozen
variables by employing for instance the replica trick, but
to keep explicitly all crosslink coordinates C during the
calculation.  Only at the very end the physical observable
of interest is evaluated for a particular realization of C
which is generated by the distribution in (\ref{4}).
Clearly both approaches will give the same results if only
self-averaging quantities are considered.

The Hamiltonian in Eq. (1) together with the uniform
distribution of crosslinks (4) defines our working model
for the ideal network.

\section{Characteristic function}

In this section a brief review of the central mathematical
theorem for the characteristic function of a Gaussian
structure with internal $\delta$-constraints is given,
together with its extension to arbitrary crosslinking
potential $z$. The characteristic function for the problem
is introduced as
\begin{equation} 
\label{5}
{\cal Z}_{0}({\bf E};\mbox{C})=\Big\langle e^{i{\bf E}\cdot
{\bf r}}
\Big\rangle_0
\end{equation}
from which all expectation values can be obtained via
differentiation. Simplifying notation has been adopted,
where ${\bf r}_j\equiv{\bf R}_j-{\bf R}_{j-1}$
$(j=1,...,N)$ denote bond vectors along the backbone of the
chain, and ${\bf E}=({\bf E}_1,..., {\bf E}_N)$, ${\bf
r}=({\bf r}_1,...,{\bf r}_N)$ are $N$-dimensional
super-vectors with $d$-dimensional vector components. Thus
${\cal Z}_{0}({\bf E};\mbox{C})$  is also the partition
function of an ideal network in the presence of external
fields ${\bf E}_j$. Note that ${\cal Z}_{0}({\bf
E};\mbox{C})$ depends explicitly on all external fields
contained in the vector ${\bf E}$, as well as on all
crosslink positions C.

Without going into mathematical details, it is now possible
to proof the following analytically {\it exact} projection
theorem \cite{solvil}
\begin{equation}
\label{6}
{\cal Z}_{0}({\bf E};\mbox{C})=\exp \left (
-\frac{a^2}{2d}\,{\bf E}_\perp^2\right ) ~,
\end{equation}
where ${\bf E}_\perp$ is the length of the external field
vector ${\bf E}$ projected perpendicular to the vector
space spanned by "crosslink vectors"
\begin{equation}
\label{7}
{\bf p}_e=(0,\dots ,0,\underbrace{1,1,\dots ,1,1}_{\mbox{
$i_e+1$ to $j_e$}},0,\dots ,0)~.
\end{equation}
The above statement can be pictured in the following
intuitive manner (figure 2). For each crosslink specified
by the pair of integers $1\leq i_e <j_e\leq N$, form the
corresponding $N$-dimensional vector ${\bf p}_{e}$, Eq.
(\ref{7}), where the 1's are assumed to run from the
$(i_e+1)$th to the $j_e$th position. The rest of the $N$
components are filled with 0's.  The whole set of $M$
vectors ${\bf p}_1,...,{\bf p}_M$ defines a characteristic
vector space for the problem, say $U$.  There is an unique
decomposition of any field vector ${\bf E}$ parallel and
perpendicular to $U$ (figure 2), viz., ${\bf E}={\bf
E}_\parallel +{\bf E}_\perp $. The operator that projects
${\bf E}$ on $U$ can be constructed from ${\bf p}_e$ for
any realization of crosslinks C. It is given by ${\cal
P}{\cal P}^+$, where ${\cal P}$ is the $N\times M$
rectangular matrix associated with the crosslink vectors
${\bf p}_e$ ($e=1,...,M$),
\begin{equation}
\label{8}
{\cal P}\equiv ({\bf p}_1,...,{\bf p}_M)~,
\end{equation}
while ${\cal P}^+$ is a generalized inverse of ${\cal P}$
\cite{linal}. 

The projection theorem (\ref{6}) is only valid for hard
crosslinking constraints $z=0$.  A generalization to
arbitrary crosslinking potential $z$ is, however, possible
using the methods of Ref. \cite{solvil}.  Only the final
result for the characteristic function is quoted which
reads
\begin{equation}
\label{9}
{\cal Z}_{0}(z,{\bf E};\mbox{C})= \exp \left[
-\frac{a^2}{2d}
\left({\bf E}^2-
\sum_{e=1}^M \frac{({\bf x}_e {\bf E})^2}
{1+(z/w_e^2)} \right )\right]~,
\end{equation}
Here ${\bf x}_e$ denotes any orthonormal basis associated
with ${\bf p}_e$ ($e=1,...,M$), and $w_e$ are corresponding
singular values \cite{linal}. For tetrafunctional
crosslinks it was shown that $w_{e}$ is always positive.
>From (\ref{9}) it is straightforward to obtain 
expressions for the radius of gyration and structure
factor.

By the scattering function (form factor) we mean density
fluctuations \cite{doied} normalized to one
\begin{eqnarray}
\label{10}
S_0({\bf k},z;\mbox{C})&=& \left\langle |\rho_{\bf k}|^2
\right\rangle_{0}\\ &\equiv &\frac{1}{N^2} \sum_{i,j=0}^N
\Big\langle\exp \Big(i{\bf k}({\bf R}_i -{\bf R}_j)\Big
)\Big\rangle_0~,\nonumber
\end{eqnarray}
where ${\bf k}$ is the scattering wavevector, and the
average is over the measure in (\ref{3}) without having
taken the $z\rightarrow 0$ limit. In the following no
distinction between $N$ and $N+1$ will be made since $N$ is
assumed to be large.  From Eq. (\ref{9}) an exact
expression for $S_{0}$ can be obtained by introducing the
external field vector ${\bf E}=k\,{\bf c}_{ij}$, where
\begin{equation}
\label{11}
{\bf c}_{ij}=(0,\dots ,0,\underbrace{1,1,\dots ,1,1}_{\mbox{
$i+1$ to $j$}},0,\dots ,0)~,
\end{equation}
and $k=|{\bf k}|$. From (\ref{9}) and (\ref{10}) it is
found
\begin{eqnarray}
\label{12}
&&S_0({\bf k},z;\mbox{C})= \frac{1}{N}\\&+& \frac{2}{N^2}
\sum_{i<j}^N \exp \left[ -\frac{a^2k^2}{2d} 
\left (j-i- \sum_{e=1}^M \frac{({\bf x}_e {\bf c}_{ij})^2}
{1+(z/w_e^2)} \right)\right]~. \nonumber
\end{eqnarray}
Similarly for the radius of gyration \cite{doied}
\begin{equation}
\label{13}
\left (\frac{R_{\mbox{\tiny g}}(z;\mbox{C})}{a}\right)^2
=\frac{N}{6}- \frac{1}{N^2}
\sum_{i<j}^N \sum_{e=1}^M \frac{({\bf x}_e {\bf c}_{ij})^2}
{1+(z/w_e^2)} ~.
\end{equation}

It is worthwhile to note that the applicability of these
results is not limited to random networks since all
crosslinking coordinates are still implicit in the formulas
through ${\bf x}_{e}$ and $w_{e}$. By selecting different
ensembles for C=$\{i_e,j_e\}_{e=1}^M$ (random or not) any
generalized Gaussian structure with internal crosslinking
constrained can be treated by the same method.  Further
generalizations of the working model (\ref{1}) to
structures built from more than one chain are feasible as
long as the objects under investigation are simply
connected. Otherwise the phantom character of the chains
and the neglect of entanglements leads to delocalization of
those clusters of the network which are not connected via
crosslinks. A similar percolation problem arises at the
vulcanization transition
\cite{zipp} which, however, is not an objective of this
study.

\section{Discussion of results}

In applying Eqs. (\ref{12}) and (\ref{13}) to random
networks, one has to deal with a sufficient number of
monomers $N$ and crosslinks $M$. Moreover, the positions of
crosslinks C=$\{i_e,j_e\}_{e=1}^M$ have to be chosen at
random. The simplest scenario  is to pick $2M$ integers
$i_e,j_e$ $(e=1,...,M)$ from the  uniform distribution, Eq.
(\ref{4}), defined on the interval $[ 0,N]$. For a given
realization C we compute an orthonormal basis ${\bf x}_{e}$
and singular values $w_{e}$ of ${\bf p}_{e}$ $(e=1,...,M)$.
Any standard technique like singular value decomposition
\cite{numrec} will suffice, for the orthonormalization
process presents only a minor numerical task. We find that
for number of monomers $N > 10000$ and crosslinks
$M > 200$ fluctuations between different realizations
of C differ by less than 1 percent.  This also presents an
estimate for the numerical uncertainty of the calculation.

\subsection{Scaling behavior of scattering function}

Within the framework of the Deam-Edwards model ($z=0$) it
was demonstrated \cite{solvil} that the scattering function
$S_{0}$ of an ideal network is an universal function of
wavevector ${\bf k}$ and mean crosslink density $M/N$ as
long as $N$ and $M$ are sufficiently large to ensure
self-averaging.  A scaling form for $S_0$ is motivated by
the following argument.  For a linear polymer without
crosslinks the scattering intensity $S_{0}(x)$ depends only
on the product $x=k^2 R_{\mbox{\tiny g}}^2$, where
$R_{\mbox{\tiny g}}^2= a^2N/6$ is the radius of gyration of
the chain and $S_0(x)=2(e^{-x}-1+x)/x^2$ the Debye function
\cite{doied}. In close analogy it was shown that for
hard constraints ($z=0$) the radius of gyration of an ideal
network is given by $R_{\mbox{\tiny g}}^2\simeq
0.26\,a^2N/M$ which suggests a scaling behavior, similar to
that of linear chains without crosslinks. This scaling
hypothesis for ideal networks was confirmed by our
calculation based on the expressions in Eqs. (\ref{12}) and
(\ref{13}). 

The numerically {\it exact} result for the scattering
function is presented in figure 3.  In the Kratky plot of
figure 3, $xS_0(x)$ has been evaluated as a function of
$\sqrt{x}\equiv kR_{\mbox{\tiny g}}$ for $z=0$.
Independent of details of crosslinking topology C, all
networks investigated fall on the same master curve (solid
line).  Statistical fluctuations between different networks
were too small in the self-averaging regime to be seen on
the scales used in figure 3.  No indication of a power-law
decay for intermediate wavevectors $k$ or other simplifying
feature was detected, besides the pronounced maximum in the
Kratky plot at $\sqrt{x}\simeq 2$ which reflects the strong
correlations of the monomers due to crosslinking.  For
comparison, the case of a linear polymer with $z\rightarrow
\infty$ (Debye function, dashed line) was also computed
from (\ref{12}).

The results of $S_{0}$ for different values of crosslinking
potential $z$ are illustrated in figure 4 for a network
with $N=10000$ and $M=200$. By increasing the strength of
the constraint from left to right, large deviations of
$S_{0}$ from ideal chain behavior (Debye function, left
curve) arise on smaller and smaller length scales. For
$z=0$ the network character persists down to even the
shortest length scale $k\approx 1$ as a
consequence of the high degree of crosslinking in the
system ($M/N=0.02$).  For sufficiently large wavevectors
all curves decay as $k^{-2}$ as expected when the scanning
wavelength becomes small compared to the mesh size of the
network. Again, the master curve for $S_{0}$ can be
obtained by plotting the $x$-axis in units of
$R_{\mbox{\tiny g}}$.

\subsection{The collapse transition}

In figure 5 we have calculated the radius of gyration of a
network of $N$ monomers and $M$ crosslinks as a function of
$\sqrt{z}=\varepsilon /a$ by use of Eq.  (\ref{13}).  From
this investigation we can clearly distinguish three different
scaling regimes
\begin{equation}
\left(\frac{R_{\mbox{\tiny g}}}{a}\right )^2 \simeq
\left \{ \begin{array}{ll}
0.26\,N/M~, & \mbox{if~} \varepsilon <<
\varepsilon_1\\ 
0.34\,(\varepsilon /a)\left(N/M\right
)^{1/2}~, &
\mbox{if~} \varepsilon_1 << \varepsilon << \varepsilon_2 \\
N/6~, & \mbox{if~} \varepsilon >> \varepsilon_2~,\\
\end{array} \right .
\end{equation}
with crossovers at $\varepsilon_1\simeq a \sqrt{N/M}$ and
$\varepsilon_2\simeq a \sqrt{MN}$. The plateau values in
figure 5 correspond to the two extremes $R_{\mbox{\tiny
g}}^2/a^2=0.26\, N/M$ ($z\rightarrow 0$) to the left and
$R_{\mbox{\tiny g}}^2/a^2 =N/6$ ($z\rightarrow \infty$) to
the right. 

In particular our investigation showed that the cases $z=0$
(hard constraints) and $z=1$ (constraints of the order of
the persistence length $a$) only differ by a numerical
prefactor which varies from 0.26 for $z=0$ to about 0.27
for $z=1$. From this we conclude that an ideal network
subject to {\it uncorrelated} crosslinking constraints is
collapsed in a sense that its size  is proportional to the
square root of $N/M$. Therefore $R_{\mbox{\tiny g}}/a\simeq
{\cal O}(1)$ since $M/N$ is the mean crosslink density in
the network and of order unity in the thermodynamic limit
$N,M\rightarrow\infty$.  This finding seems to be at
variance with current speculations regarding the collapse
transition of macromolecules \cite{bry}, where it was
argued that a critical number of crosslinks $M\geq
M_c\simeq N/\log N$ will force the system to collapse. Our
result for $z=0$ is in agreement with recent Monte Carlo
simulations by Kantor and Kardar \cite{kant} who found for
the mean squared  end-to-end distance $R^2/a^2\simeq
1.5\,N/M$.  This suggests the same one to six ratio for
$(R_{\mbox{\tiny g}}/R)^2$ in ideal networks as for linear
polymers without crosslinking constraints and
excluded-volume interaction \cite{doied}.
We believe that the discrepancy between our exact results
and these suggested in \cite{bry} are due to the
approximations used there. 

Conversely, a free chain ($z\rightarrow \infty$) is an
extended object with $R_{\mbox{\tiny g}}/a\simeq{\cal
O}(\sqrt{N})$. Between the collapsed and the extended
regime we find a smooth crossover with $(R_{\mbox{\tiny
g}}/a)^2$  being proportional to $(\varepsilon
/a)\sqrt{N/M}$.  Remarkably this is  the
same scaling as for randomly branched polymers without
excluded volume interaction. In the following we discuss
the different conformational states of the system in terms
of simple scaling arguments.

\subsection{Flory estimates}

For completeness we first note that the free chain regime
is trivial since $R_{\mbox{\tiny g}}^2/a^2 =N/6$ is an
exact solution of (13) for $z\rightarrow\infty$.  To understand
the scaling behavior of $R_{\mbox{\tiny g}}$ in the other 
two regimes
the free energy of the Hamiltonian in Eq.  (1) is discussed
within Flory theory.  The connectivity term in (1) models
the standard entropic elasticity of a Gaussian chain, i.e.,
$R^2/(Na^2)+N a^2/R^2$, where $R$ is a measure of the size
of the system.  The first term accounts for  stretching,
whereas the second term describes the response due to
compression \cite{deGennes}.

An estimate of the crosslink term in (1) requires more
attention.  First we consider soft crosslinks when
$\varepsilon >> a$.  In this regime the second term of the
Hamiltonian is estimated by $M (R/\varepsilon )^2$,
because the mean squared distance between a pair of constrained
monomers is of order $\varepsilon^2$. The relevant part of
the total Flory free energy then is
\begin{equation}
{\cal F}_{0} \sim \frac{Na^2}{R^2}+\frac{M R^2}
{\varepsilon^2}~.
\end{equation}
Minimization of the free energy with respect to $R$ yields
the scaling relation $R/a \sim (\varepsilon/a)^{1/2}
(N/M)^{1/4}$ in agreement with (14).  The appearence of the
branched polymer exponent $1/4$ can be assigned to the
change of connectivity of the chain when 
$\varepsilon \simeq {\cal O}(\sqrt{NM})$.

The case of hard crosslinks $\varepsilon \simeq {\cal
O}(a)$ is more difficult to obtain.  We picture the system
as a coarse-grained random walk over the $M$ crosslinks
with an effective step length proportional to $N/M$, i.e.,
the mean number of monomers between crosslinks. From this
mean-field argument the crosslink term is estimated to be
of the order $M [R^2/(a^2 N/M)]$. The latter expression has
the effect that it tries to shrink the chain upon cost of
confinement entropy. A suitable Flory free energy is given
by
\begin{equation}
\label{flory}
{\cal F}_{0} \sim \frac{N a^2}{R^2} + \frac{M^2 R^2}{Na^2}~.
\end{equation} 
>From there $(R/a) \sim (N/M)^{1/2}$ as was shown exactly for 
the network regime.

\acknowledgements
M.P.S. gratefully acknowledges financial support by the
DFG, Sonderforschungsbereich 262.


\begin{figure}
\caption[x]{Modelling of a soft crosslink
between two monomers $i_e$ and $j_e$. A spring constant
$\varepsilon\rightarrow 0$ leads to hard
$\delta$-constraints, $\varepsilon\rightarrow \infty$ to
free chain behavior.}
\label{fig1}
\end{figure}

\begin{figure}
\caption[x]{External field vector ${\bf E}$ projected
parallel and per\-pen\-dicular to $U$. The vector space $U$
is spanned by the vectors ${\bf p}_e$ defined in  Eq. (7).
The sketch is for two crosslinks $e=1,2$.}
\label{fig2}
\end{figure}

\begin{figure}
\caption[x]{Kratky plot of $x S_{0}(x)$ for the ideal
network ($z=0$, solid line) and linear chain ($z\rightarrow
\infty$, Debye function, dashed line) as a function of
$\sqrt{x}=kR_{\mbox{\tiny g}}$ in $d=3$. The open circles
show a second realization of C to demonstrate
self-averaging. Note that we measure length scales in terms
of $R_{\mbox{\tiny g}}$ and use a normalization
$S_{0}(0)=1$ different to that in the experimental literature.}
\label{fig3}
\end{figure}

\begin{figure}
\caption[x]{Structure function $S_{0}$ for different
crosslinking potentials $z$. From right to left
$z=0,10,10^2,10^3,10^4,10^5,\infty$ for a network size
of $N=10000$ and $M=200$ plotted in dimensionless units
$q=ka/\sqrt{2d}$. The left curve ($z\rightarrow\infty$) is
the Debye function of a linear chain.}
\label{fig4}
\end{figure}

\begin{figure}
\caption[x]{Radius of gyration of a polymer chain as a
function of $\sqrt{z}= \varepsilon /a$. For the three solid
curves the number of monomers was varied from top to bottom
$N=5000,10000,20000$; $M$ was kept constant at 200. The
short dashed line shows a network with $N=20000$ and
$M=400$.}
\label{fig5}
\end{figure}

\end{document}